\documentclass[preprint,12pt,3p]{elsarticle}

\usepackage{multirow}
\usepackage{amssymb}
\usepackage{comment}
\usepackage{amsfonts}
\usepackage{amsmath}
\usepackage{siunitx}
\usepackage{amsthm}
\usepackage{eucal}
\usepackage{color,soul}
\usepackage{mathrsfs}
\usepackage{lineno}
\usepackage{graphicx}
\usepackage{epstopdf}
\usepackage{caption}
\usepackage{nomencl}
\usepackage[toc,page]{appendix}
\usepackage{subcaption}

\usepackage{hyperref} 
\hypersetup{colorlinks=true,linkcolor=blue,hyperfootnotes=false,citecolor=blue}

\begin{document}

\begin{frontmatter}

\title{Intelligent mechanical metamaterials towards learning static and dynamic behaviors}

\author[add1]{Jiaji Chen}
\author[add1]{Xuanbo Miao}
\author[add1]{Hongbin Ma}
\author[add2]{Jonathan B. Hopkins}

\cortext[mycorrespondingauthor]{Corresponding author}

\author[add1]{Guoliang Huang\corref{mycorrespondingauthor}}
\ead{huangg@missouri.edu}

\address[add1]{Department of Mechanical and Aerospace Engineering, University of Missouri, Columbia, MO 65211, USA}

\address[add2]{Mechanical and Aerospace Engineering, University of California, Los Angeles, CA 90095, USA}


\begin{abstract}

The exploration of intelligent machines has recently spurred the development of physical neural networks, a class of intelligent metamaterials capable of learning, whether \textit{in silico} or \textit{in situ}, from observed data. In this study, we introduce a back-propagation framework for lattice-based mechanical neural networks (MNNs) to achieve prescribed static and dynamic performance. This approach leverages the steady states of nodes for back-propagation, efficiently updating the learning degrees of freedom without prior knowledge of input loading. One-dimensional MNNs, trained with back-propagation \textit{in silico}, can exhibit the desired behaviors on demand function as intelligent mechanical machines. The framework is then employed for the precise morphing control of the two-dimensional MNNs subjected to different static loads. Moreover, the intelligent MNNs are trained to execute classical machine learning tasks such as regression to tackle various deformation control tasks. Finally, the disordered MNNs are constructed and trained to demonstrate pre-programmed wave bandgap control ability, illustrating the versatility of the proposed approach as a platform for physical learning. Our approach presents an efficient pathway for the design of intelligent mechanical metamaterials for a wide range of static and dynamic target functionalities, positioning them as powerful engines for physical learning. 

\end{abstract}

\end{frontmatter}

\section{Introduction}
Mechanical metamaterials can obtain distinct and exotic mechanical properties and behaviors through the rational design of their microstructures in periodic and disordered fashions \cite{wu2022independent, wu2023active, chen2023broadband}. However, passive metamaterials can not adapt their responses and functionalities to external loadings. To overcome this challenge, intelligent mechanical metamaterials were suggested to tune their mechanical responses by integrating smart or responsive materials with sensing and information processing into their microstructures and creating a sense-decide-response loop \cite{le2019filtered, riley2022neuromorphic, liu2023discriminative}. Intelligent mechanical metamaterials have been successfully employed for tunable static and dynamic control behaviors including real-time deformation, vibration and wave propagation control \cite{chen2018programmable, zhang2023embodying, ren2021smp, li2014granular, li2018self, chen2016adaptive}. In those applications, the global properties of the intelligent metamataerials were typically adjusted for expected loading conditions to adjust their functionality and performance. However, designing intelligent mechanical metamaterials for auto-adaptive and multiple task purposes is very expensive and cumbersome by reevaluating their global properties offline \cite{el2021digital}. How to design intelligent metamaterials with learning abilities as mechanical machines with time-varying stiffness to on-line realize the sought-after mechanical behavior for any unexpected loadings is still very challenging. 

Inspired by the remarkable success of machine learning (ML) \cite{cambria2014jumping, rawat2017deep, silver2016mastering, gilmer2017neural, torlai2018neural, chen2022physics, fu2022identifying, hinton2012deep, hirschberg2015advances}, recent advancements led to the development of physical neural networks (PNNs), a new class of intelligent metamaterial capable of learning desired behaviors online. Basically, there are two principle types of PNN. The first type is structured analogously to artificial neural networks (ANN), featuring layers of metasurfaces or coupled resonators that simulate the roles of neurons and synapses \cite{weng2020meta, lin2018all, qu2022resonance, jiang2023metamaterial}. These structures compute outputs as complex functions of inputs and tunable parameters, effectively utilizing the same principles as digital computation systems. However, the practical engineering applications of these PNNs have been limited to computational tasks, which do not directly translate into tangible mechanical functionalities or adaptive material properties required in real-world engineering applications \cite{hamerly2019large, pankov2022optical, yan2019fourier, zuo2021scalability}. Alternatively, the second type of PNN that diverges significantly from the conventional digital neural network architectures has shown promise in engineering utilities. Pashine \textit{et al.} demonstrated a disordered elastic network with a negative Poisson ratio \cite{pashine2019directed}. Similarly, Tang \textit{et al.} designed a self-activated solid for uni-mode and bi-mode extremal materials \cite{tang2024learning}. These systems adaptively respond to external inputs by altering their physics degrees of freedom to maintain structural balance, such as adjusting node positions in elastic networks or modifying current distributions in accordance with Kirchhoff's laws in electrical networks \cite{stern2021supervised, stern2023learning}. The physics degrees of freedom are intimately coupled with the learning degrees of freedom in PNNs, which are the adjustable parameters of the system (like stiffness in elastic networks or edge conductance in electrical networks). This type of PNN can achieve complex input-output mapping without replicating the structure of ANN \cite{wright2022deep}. The primary design challenge with these systems is the computation of gradients necessary for effective back-propagation, as their unique physical architectures do not readily lend themselves to conventional back-propagation techniques. Nevertheless, in design of lattice-based intelligent mechanical metamaterial, called mechanical neural networks (MNNs), where learning processes are slower than physical response rates, the gradients can be effectively determined by exploiting the force equilibrium at each node.

\begin{figure}[ht!]
\includegraphics[width=\linewidth]{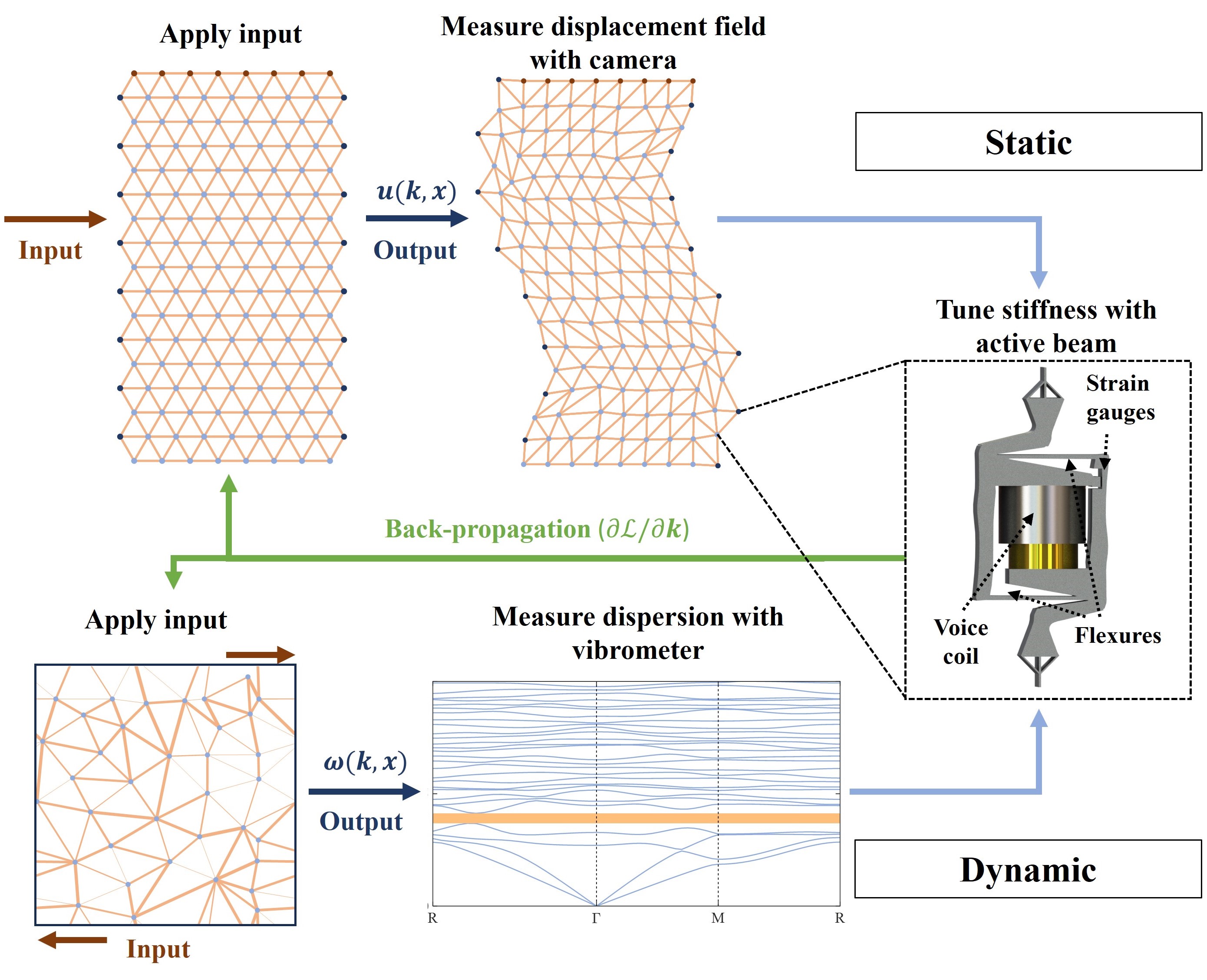}
\caption{\label{fig1} Architecture visualization of a mechanical neural network. The top panels show the training loop of a static lattice, while the bottom panels show that of a disordered dynamic lattice. The edge stiffness $\pmb{k}$ is updated with back-propagation $\partial \mathcal{L} / \partial \pmb{k}$. $\mathcal{L}$ denotes the loss function. A design of edge with tunable stiffness is shown in the right panel. The design consists an active beam that achieve tunable stiffness through closed-loop active control, utilizing voice coils and strain gauges to detect and actuate flexure deformations, guiding the extension and contraction of the beams along their axes.}
\end{figure}

This study introduces a novel back-propagation framework for lattice-based MNNs, which comprise hinged active beams with adjustable stiffness, as shown in Fig. \ref{fig1}. This lattice respond to unknown force or displacement inputs, with its node movements captured by external sensors like high-speed cameras and laser vibrometers in static and dynamic settings, respectively \cite{lee2022mechanical}. We utilize implicit differentiation at force-balanced nodes to compute the gradients of the stiffness, facilitating precise control over the lattice's response. The intricate coupling between the lattice edges and nodes allows for complex deformations, which feature geometric nonlinearity. By integrating back-propagation, the mechanical metamaterial is enabled to learn, morph, and adapt into specific shapes under various loading or displacement conditions. One-dimensional MNNs, trained with back-propagation \textit{in silico}, can exhibit the desired behaviors on demand function as intelligent mechanical machines. The approach is then employed for the precise morphing control of the two-dimensional MNNs subjected to shear or uniaxial loads. Moreover, the intelligent MNNs are trained to execute classical machine learning tasks such as regression. To illustrate the versatility of the approach, the desired bandgap control ability for the disordered lattice is also effectively tuned. Our proposed back-propagation framework opens new possibilities for the inverse design of MNNs and other intelligent metamaterials, targeting any specified properties or functionalities.

\section{Prescribed deformation control for one-dimensional MNN}

We first present how the back-propagation is employed in the context of a mechanical lattice, shown in Fig. \ref{fig1}. The stiffness of the edges are tuned as learning degrees of freedom to train the lattice such that it can learn target mechanical behaviors without prior knowledge about given input. The node positions serve as physical degrees of freedom and respond differently to different stiffness and input boundary conditions. 

\subsection{back-propagation in ANN}
In an ANN with $L+1$ layers (including input and output layers), forward propagation maps the neuron activations from layer $l-1$ to neuron inputs at layer $l$ as $z_j^{(l)}=\sum_{i} w_{ji}^{(l)} a_i^{(l-1)}$ via a weight matrix $w^{(l)}$, before applying a nonlinear activation function individually to each neuron, $a_j^{(l)}=g(z_j^{(l)})$. Dummy variables $i$ and $j$ are the indices for neurons in the current layer. At the output layer, we evaluate the loss function, $\mathcal{L}$, and calculate its gradient with respect to the weights \cite{guo2021backpropagation}:
\begin{equation}
\label{eq1}
\frac{\partial \mathcal{L}}{\partial w_{ji}^{(l)}}=\frac{\partial \mathcal{L}}{\partial z_{j}^{(l)}} \frac{\partial z_{j}^{(l)}}{\partial w_{ji}^{(l)}}=\delta_j^{(l)} a_i^{(l-1)}
\end{equation}
where $\delta_j^{(l)}=\partial \mathcal{L} / \partial z_{j}^{(l)}$. From the chain rule, we have 
\begin{equation}
\label{eq2}
\delta_j^{(l)}=\sum_k \frac{\partial \mathcal{L}}{\partial z_{k}^{(l+1)}} \frac{\partial z_{k}^{(l+1)}}{\partial z_{j}^{(l)}}=g'(z_{j}^{(l)}) \sum_k w_{kj}^{(l+1)} \delta_k^{(l+1)}
\end{equation}
Given the error at the output layer $\delta^{(L)}$, which is calculated directly from the loss function, $\delta^{(L-1)}$, ..., $\delta^{(1)}$ for all preceding layers are sequentially determined using Eq. \ref{eq2}, and the iterative process is referred as the back-propagation. 

For the back-propagation of an static lattice considered in Fig. \ref{fig1}, the learning problem is to accomplish the target displacement output at the left and right boundary nodes under a displacement or force load applied to the top nodes. Similar equations to Eq. \ref{eq2} are not applicable in this case given the distinct architectural differences between MNNs and ANNs. Alternative training methods have been developed to address this issue. For instance, Lee \textit{et al.} used pseudo-gradient method such as Nelder-Mead algorithm to update learning degrees of freedom \cite{lee2023comparing}. Stern \textit{et al.} compared outputs of free state and clamped state to modify learning degrees of freedom \cite{stern2021supervised}. Despite efforts to develop new learning rules, resulting designs require either intensive computational resources or exhaustive experimental conditions.

\begin{figure}[ht!]
\includegraphics[width=0.7\linewidth]{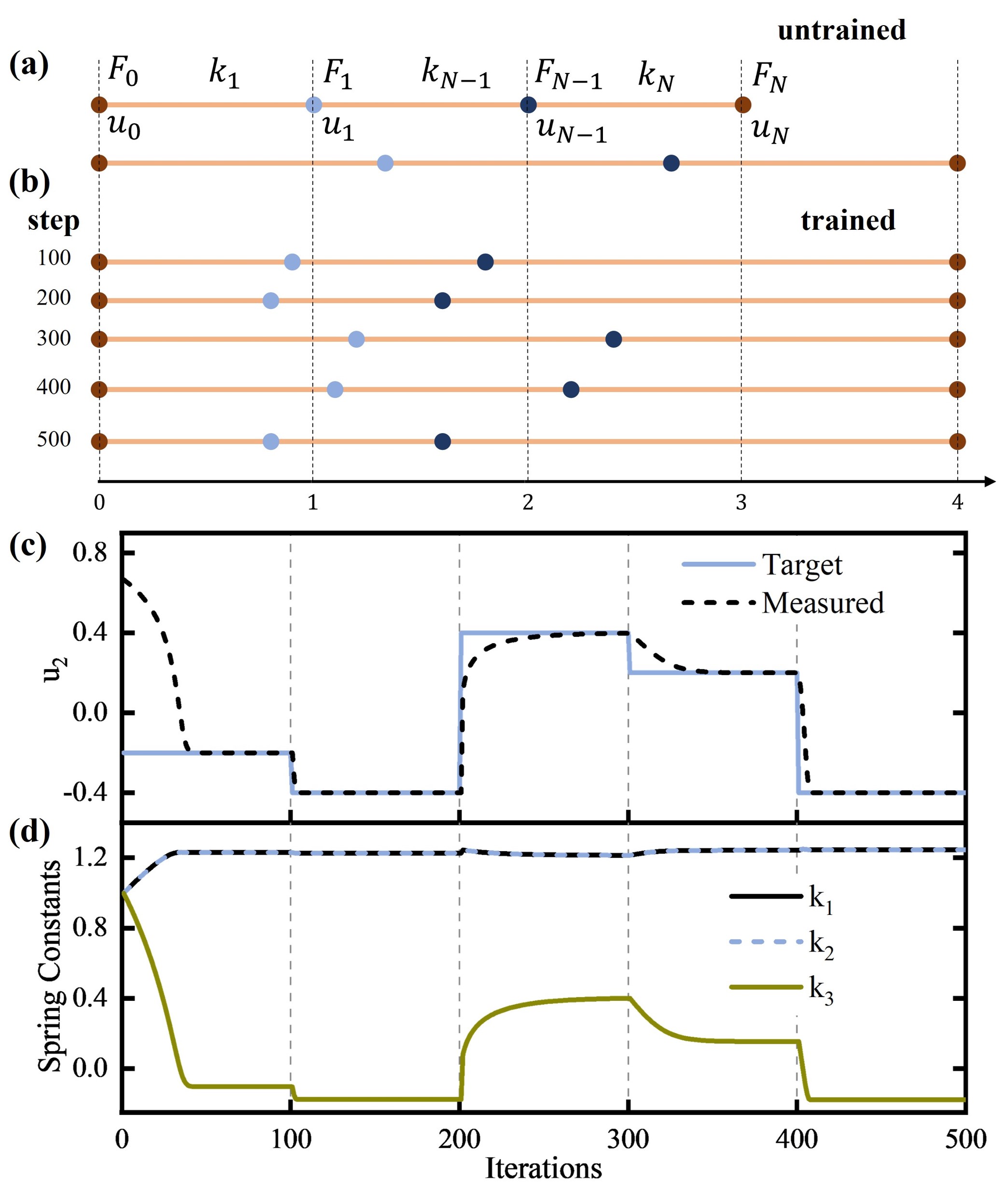}
\centering
\caption{\label{fig2} (a) Schematic illustration of relaxed one-dimensional MNN of $N$ edges (top). When the lattice is under tension, its left node is fixed, and the right node has displacement $u_N = 1$. (b) After the lattice is trained to targeted functionality, its deformation at each step is shown. (c) The target displacement $u^{target}$, and the output node (node 2, dark blue) displacement $u_2$ versus training steps. (d) The stiffness of the three edges as a function of training steps. }
\end{figure}

\subsection{back-propagation in MNN}
To gain access to $\partial \mathcal{L} / \partial \pmb{k}$, which is necessary for evaluating the updates of each edge stiffness, we use implicit differentiation of output with respect to stiffness. Considering the simplest case, where our metamaterial is a one-dimensional lattice of N edges, as shown in Fig. \ref{fig2}(a). Each edge is deformed due to the nearest-neighbor interaction. Displacement boundary conditions are given to left and right boundary nodes. Since the lattice is in equilibrium, the net force on each node is zero. For the $n$-th node, we get:
\begin{equation}
\label{eq3}
F_n = k_n (u_n - u_{n-1}) + k_{n+1}(u_n-u_{n+1})=0
\end{equation}
where $k_n$ and $u_n$ are stiffness of $n$-th edge and displacement of $n$-th node, respectively. For $n=0$ or $N$, we have $F_0 = F_l + k_1 (u_0 - u_1) = 0$ and $F_N = k_N (u_N - u_{N-1}) + F_r = 0 $, where $F_{l/r}$ is the boundary force on the left/right node. 

To find gradient $\partial u^{out}/\partial k_{m}$, $m=1,2...N$, where $u^{out}$ is the displacement output for the defined output nodes, we can solve the following equation 
\begin{eqnarray}
\label{eq6}
\sum_{j=0}^{N} \frac{\partial F_i}{\partial u_j} \frac{\partial u_j}{\partial k_m} = \frac{\partial F_i}{\partial k_m}, \qquad
\begin{array}{c}
i,j = 0,1,\cdots,N\\
m = 1,2,\cdots,N
\end{array}
\end{eqnarray}
where $\partial F_i/\partial u_j$ and $\partial F_i/\partial k_m$ can be derived from Eq. \ref{eq3}, and $\partial u_j/\partial k_m$ is given by Eq. \ref{eq6}. Thus, the update rule for stiffness can be established as:
\begin{eqnarray}
\label{eq7}
\Delta k_m  = - \eta \frac{\partial \mathcal{L}}{\partial k_m} = \sum_{i=0}^{N} - \eta \frac{\partial \mathcal{L}}{\partial u_i} \frac{\partial u_i}{\partial k_m}
\end{eqnarray}
where $\eta$ is user defined learning rate and $\mathcal{L}$ denotes the loss function. Unless specified, we use $\eta = 0.01$ and mean squared error (MSE) as a loss function throughout this paper.  To train a lattice to exhibit target behaviors, we need to first define input nodes, which is essentially equivalent to assigning boundary conditions to the network. We adjust $u^{out}$ to align with the target pattern under the given boundary conditions. To accomplish that, we iteratively apply $\Delta k_m$ to the stiffness for a defined period of time. Similar principles apply to dynamic lattice, and their details are in section \ref{dynamic}.

To demonstrate the power of our proposed method, training is conducted on a one-dimensional lattice of $N=3$ edges under tension or displacement loads. For example, the objective is to guide the displacement of node 2 (dark blue node in Fig. \ref{fig2}) toward $u^{target}$, while maintaining displacements of node 0 and node 3 fixed at 0 and $u_3=1$, respectively. Each spring has a rest length of $L_0 = 1$, and the initial spring constant is set to be $k_m = 1$ for all edges. Here,  $u^{target}$ is changed for every 100 training steps. As indicated in the solid line in Fig. \ref{fig2}(c), $u^{target}=-0.2,-0.4,0.4,0.2,-0.4$, respectively. As the training progress, $u^{out}$ rapidly converge to $u^{target}$ in 30-40 iterations. Similar patterns can be observed as $u^{target}$ changes. Notice that in iteration 0-100, 100-200, and 400-500, $u^{target}$ is negative. To achieve negative displacement under the given boundary condition, the effective spring constant of $k_1$ and $k_2$, which is $k_1 k_2/(k_1 + k_2)$, must have the opposite sign to $k_3$. We can observe from Fig. \ref{fig2}(d) that $k_1$ and $k_2$ stay positive during the training, and so does their effective one, while $k_3$ goes negative in the intervals 0-100, 100-200, and 400-500. The trained mechanical lattices of different $u^{target}$ can be found in Fig. \ref{fig2}(b). 

\begin{figure}[ht!]
\includegraphics[width=\linewidth]{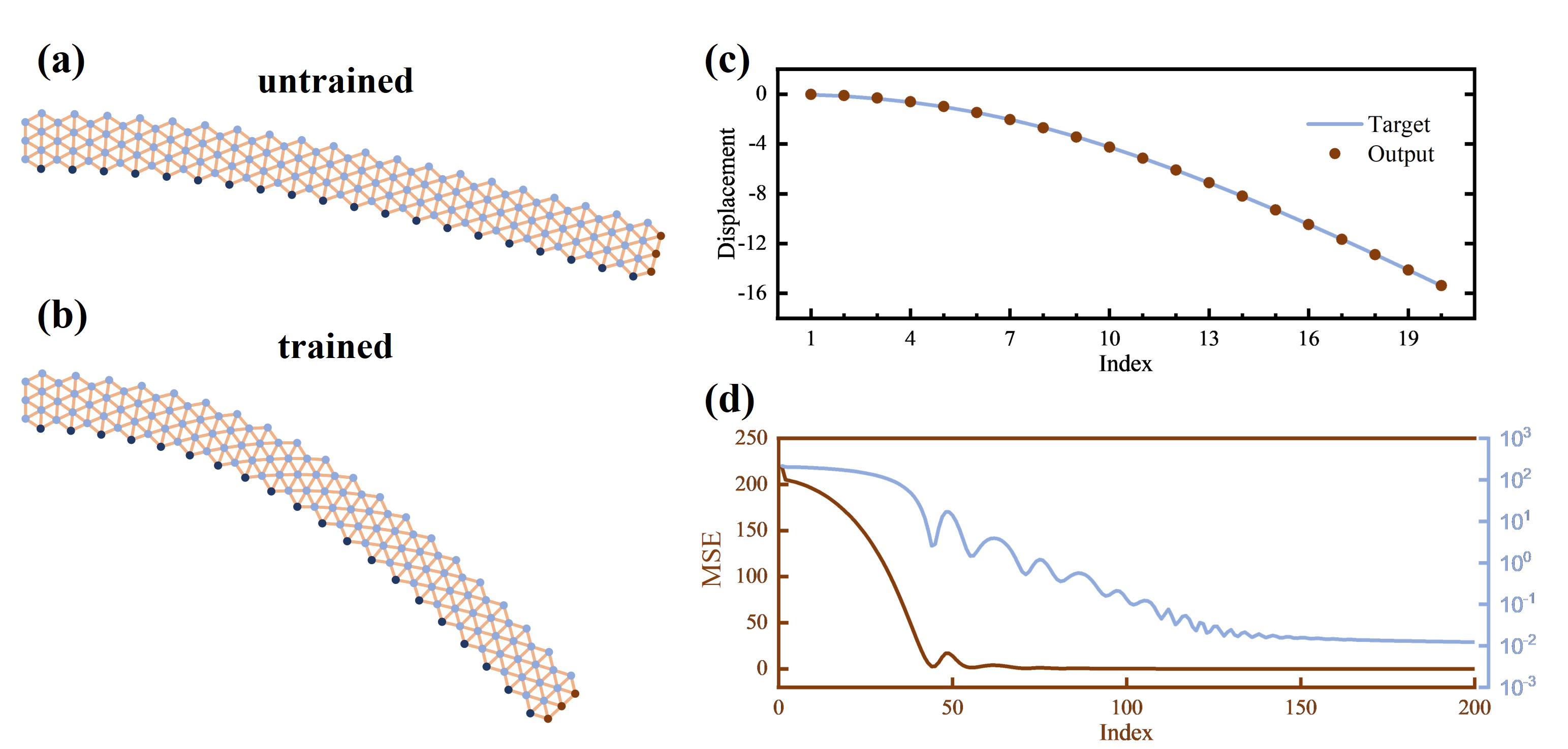}
\centering
\caption{\label{sup_fig1} Training MNN under shear stress (a) Deformation of an untrained MNN with uniform edge stiffness $k_m=1$. The 3 red nodes on the right are input nodes under the loading boundary condition of $F^y = -0.001$. The 3 nodes on the left are fixed. The 20 dark blue nodes on the bottom are output nodes. (b) Deformation of a trained MNN. The thickness of the edges is normalized with the maximum stiffness. (c) The target displacement $u^{target}$, and the output node (dark blue) displacement $u^{output}$. (d) The loss function versus training steps. The light blue curve corresponds to the loss function plot in the logarithmic scale. } 
\end{figure}

\section{Morphing control for two-dimensional MNN}
We now demonstrate the capability of our back-propagation framework by training two-dimensional lattices as intelligent material. The learning rule of two-dimensional MNN is similar to that of one-dimensional one except for the nonlinearity in Euclidean distance function. For a lattice of $N$ nodes and $M$ edges, the force equilibrium at i-th node is established as:
\begin{equation}
\label{eq8}
\pmb{F}_i = \sum_{j=1}^{T} k_{ij}(dL_{ij}-L_0) \frac{\pmb{dx}_{ij}}{dL_{ij}} = \pmb{0}
\end{equation}
where $\pmb{F}_i = [F_i^x, F_i^y]$ is the force acting on $i$-th node with the superscript $x$ and $y$ being the $x$ and $y$ components of the vector, $T$ is the number of nodes connected to the i-th node with edges, $k_{ij}$ is the stiffness of the edge between i-th and j-th node, $L_0$ is the rest length of edges, $\pmb{dx}_{ij} = \pmb{x}_{j} - \pmb{x}_{i}$ is the coordinate of j-th node from i-th node, and $dL_{ij}=|\pmb{dx}_{ij}|$ denotes the Euclidean distance between i-th node and j-th node. The corresponding implicit differentiation for solving the gradient of displacement for the spring constant is formulated as:
\begin{eqnarray}
\label{eq11}
\sum_{j=1}^{N} \frac{\partial \pmb{F}_i}{\partial \pmb{u}_j} \frac{\partial \pmb{u}_j}{\partial k_m} = \frac{\partial \pmb{F}_i}{\partial k_m}, \qquad
\begin{array}{c}
i,j = 1,2,\cdots,N\\
m = 1,2,\cdots,M\\
\end{array}
\end{eqnarray}
and $\pmb{u}_i = [u_i^x, u_i^y]$ denotes the node displacement.

\begin{figure}[ht!]
\includegraphics[width=\linewidth]{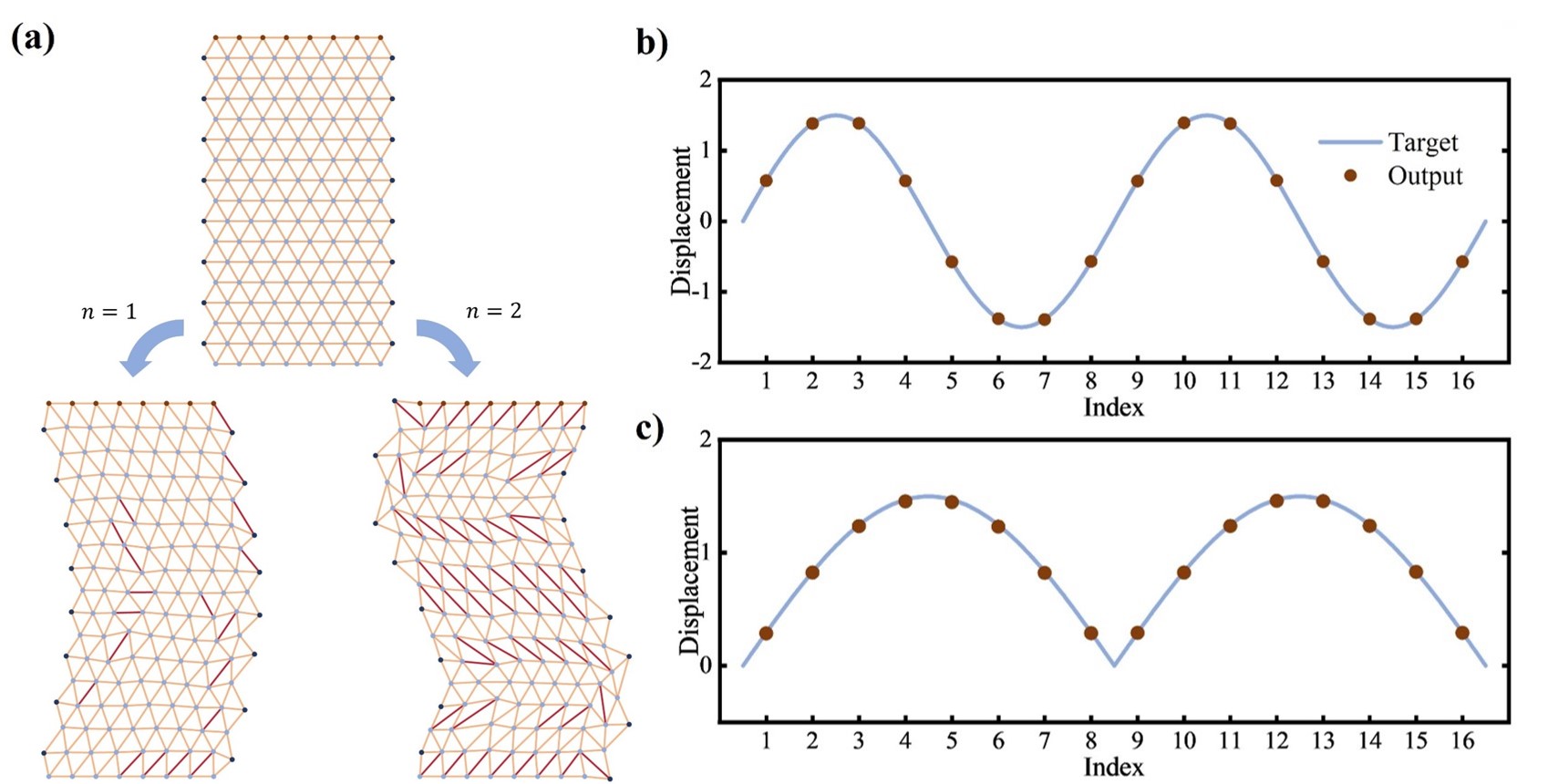}
\caption{\label{fig3} Training MNN under displacement boundary condition (a) (Top panel) Deformation of an untrained MNN with uniform edge stiffness $k_m=1$. The red nodes on top are input nodes under the displacement boundary condition of $u_i^y = 0.3$. The nodes on the bottom are fixed. The dark blue nodes on the left and right are output nodes. (Bottom panels) Deformation of trained MNNs. The grey edges represent those with negative stiffness. (b) The target displacement $u^{target}$, and the output displacement $u^{output}$ when $n=2$. Indices 1-8 correspond to nodes on the left boundary, and indices 9-16 correspond to the right one. (c) The target displacement $u^{target}$, and the output displacement $u^{output}$ when $n=1$.} 
\end{figure}

Here, a triangular lattice as a 2D intelligent metamaterial is suggested and trained, as shown in Fig. \ref{sup_fig1}(a). The left nodes of the lattice are fixed, and the right nodes are under loading condition $F^y = -0.001$. The lattice is trained so that its bottom nodes displacement ($u_i^x$, $u_i^y$) follows $u_i^y = A_0 (\cos(\pi x^i/l_0) - 1)$, where $x^i$ is the $x$ coordinate of output nodes before applying boundary conditions, $l_0$ denotes the length of the bottom boundary and $A_0 = 16$ is the amplitude of the deformation. Fig. \ref{sup_fig1}(c) shows that trained lattice has output node displacement $u^{output}$ that fits perfectly well with $u^{target}$. The MNN succeeds in achieving the desired deformation shape, reducing the MSE by orders of magnitude (Fig. \ref{sup_fig1}(d)). The thickness in the figure is proportional to the spring constants. It is obvious that the trained lattice has a much larger curvature compared with the untrained one under the same loads.  One can observe that edges near the bottom boundary possess smaller spring constants compared with those on the top boundary. Thus the lattice is pruned to bend more and fit into our target shape. This suggests that the lattice succeeded in transforming from the original configuration (Fig. \ref{sup_fig1}(a)) into the trained state (Fig. \ref{sup_fig1}(a)) under large deformation. Note that we did not tune hyperparameters to achieve such performance. Further tuning hyperparameters could improve the performance of our designs.

Alternatively, a bulk 2D MNN is trained for prescribed shape morphing on side boundaries. As depicted in Fig. \ref{fig3}, a bulk triangular lattice is demonstrated to possess the capability of learning two distinct sinusoidal shape-morphing behaviors prescribed at its boundaries. The sinusoidal pattern of the output displacement ($u_i^x$, $u_i^y$) is defined as $u_i^x = A_0 \sin(n \pi y^i/l_0)$, with a prescribed amplitude of $A_0 = 1.5$. In this setup, the lattice's bottom nodes are fixed, while the top nodes serve as input nodes subjected to a displacement condition of $u_i^y = 0.3$. The left and right nodes are designated as output nodes. A clear sinusoidal shape can be observed in the trained lattices (bottom panels of Fig. \ref{fig3}(a)). There is a good agreement between the desired shape (solid line) and the trained morphing shape (dots), as illustrated in Fig. \ref{fig3}(b,c). Notice that it is difficult for a conventional lattice (with all edge stiffness positive) to exhibit such behavior. When the amplitude of the target sinusoidal shape is large, part of the lattice expands under stretching. This requires auxetic materials that have a negative Poisson ratio. Conventional triangular lattices, having positive Poisson's ratios, are unsuitable for this task. Therefore, in our model, the edge stiffness is not limited to positive values, and it may also include negative values to meet specific requirements. Given that all edge stiffness were initially set to 1, there is an inherent bias towards positive values in the lattice. Consequently, the resulting lattice comprises 95.04\% positive edges for $n=1$ and 85.90\% for $n=2$. This study underscores the efficacy of our proposed mechanical metamaterial, which can rapidly and accurately learn and adapt to various shapes. Observing from the trained lattice of $n=1$, we can find that most of the edges with negative stiffness are located at the bottom boundary, right boundary, and slightly left to the middle. The edges on the right boundary are merely for fine-tuning the shape. What is causing the lattice to morph into a sinusoidal shape is because of those negative edges on the bottom and in the middle. When the lattice is stretched, horizontal edges in the middle tend to experience compression force. Negative springs extend with compression and push the body of the lattice toward the right, resulting in the current shape morphing. Similar phenomena can be found in the case $n=2$.

\begin{figure}[ht!]
\includegraphics[width=0.8\linewidth]{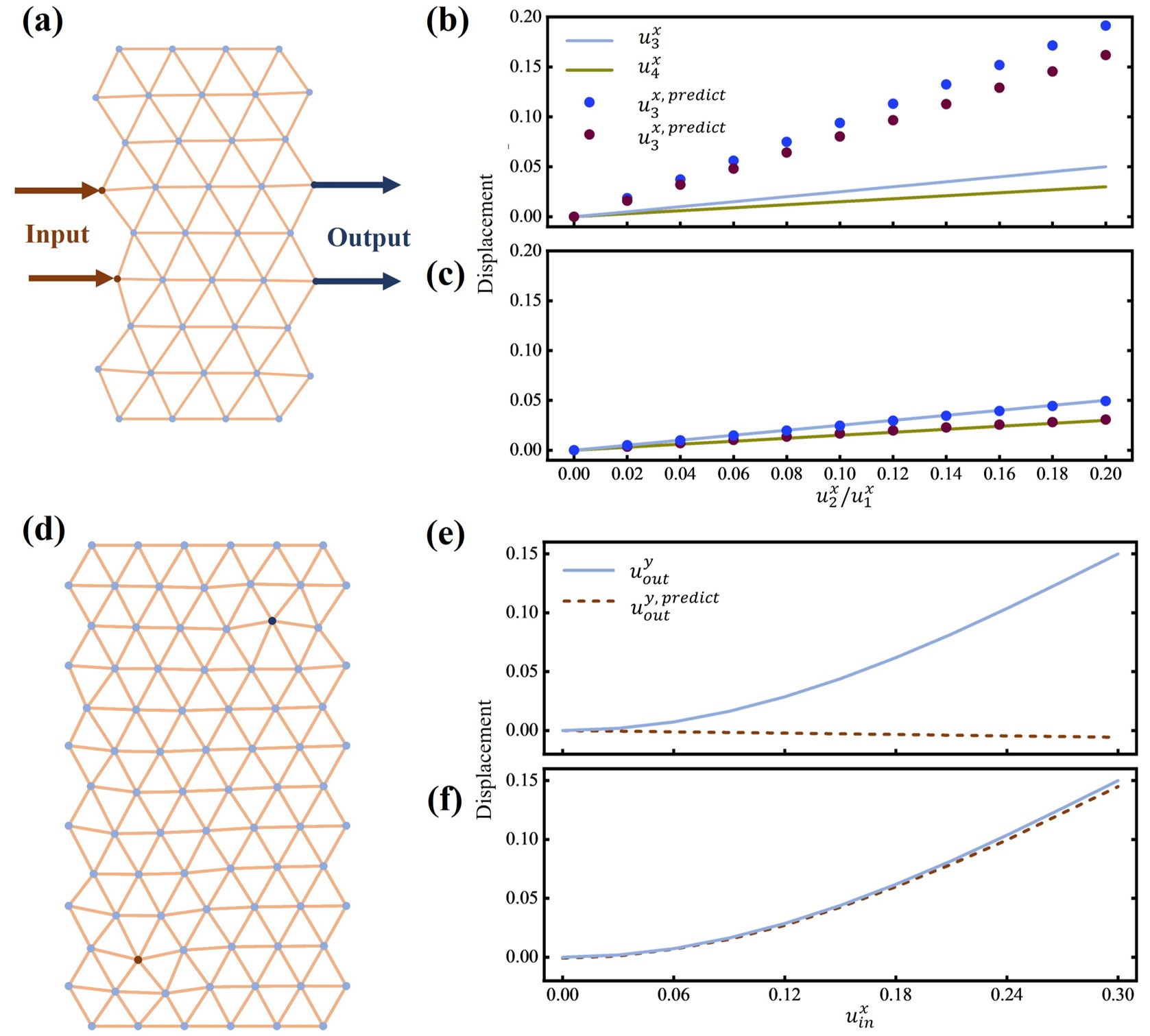}
\centering
\caption{\label{fig4} Training MNN for regression task (a) MNN setup for training a regression network to fit Eq. \ref{eq12} (b) The plot for both outputs before training. The solid lines indicate the desired output values. Since regression involves two parameters, both axes are scaled by $u_1^x$ to project the results into two dimensions. (d) The plot for both outputs after training. } 
\end{figure}

The proposed back-propagation framework can be further customized to tackle various task types, presenting a more challenging assessment. Specifically, the metamaterial is requested to perform a prescribed response behavior, which is presented in the following two equations:
\begin{subequations}
\label{eq12}
\begin{equation}
u_3^{x} = 0.15u_1^{x} + 0.20 u_2^{x}
\end{equation}
\begin{equation}
u_4^{x} = 0.25u_1^{x} + 0.10 u_2^{x}
\end{equation}
\end{subequations}
Here $u_1^{x}$ and $u_2^{x}$ denote the displacement in the x-direction of 2 input nodes on the left boundary, and $u_3^{x}$ and $u_4^{x}$ denote the displacement in the x-direction of 2 output nodes on the right boundary (Fig. \ref{fig4}(a)). A dataset of 100 is randomly generated by choosing input pair values between 0 and 0.3 and calculating the desired voltage for each input pair using the above equations. The data is then divided into a training set of 150 pairs and a testing set of 50. The untrained lattice output deviates from the target at initialized status, whereas the trained one gives accurate prediction on the regression task (Fig. \ref{fig4}(b,c)). A similar principle applies to nonlinear input-output relations. In Fig. \ref{fig4}(d), a node is randomly picked as the input node, and another one is selected as the output. We then ask the $u_{out}^{y} = 0.15 (\cos(\pi u_{in}^{x} / 0.3)-1)$. The trained lattice's behavior is in good agreement with the desired response ( Fig. \ref{fig4}(f)). The proposed learning rules are generally able to train the networks to exhibit the desired responses, successfully decreasing the initial error by orders of magnitude. 

The accuracy with which a mechanical lattice learns specific behaviors is intrinsically tied to the size of the network. This concept mirrors the principle observed in neural networks, where a network that is both deeper and wider possesses a greater number of parameters within its function. Such an increase in parameters results in a more sophisticated mapping from input to output, enabling the network to discern and replicate more complex patterns within the data. This principle is equally applicable to mechanical networks. Their ability to represent and learn is enhanced by augmenting either the learning degrees of freedom, through the addition of more edges, or the physical degrees of freedom, through the addition of more nodes. Consequently, the presence of a larger number of nodes and edges directly correlates with the network’s ability to learn and adapt to shape deformations. By continuously increasing the number of nodes and edges, one can progressively refine the accuracy with which the network learns and adapts to shape deformations. 

\begin{figure}[ht!]
\includegraphics[width=17cm]{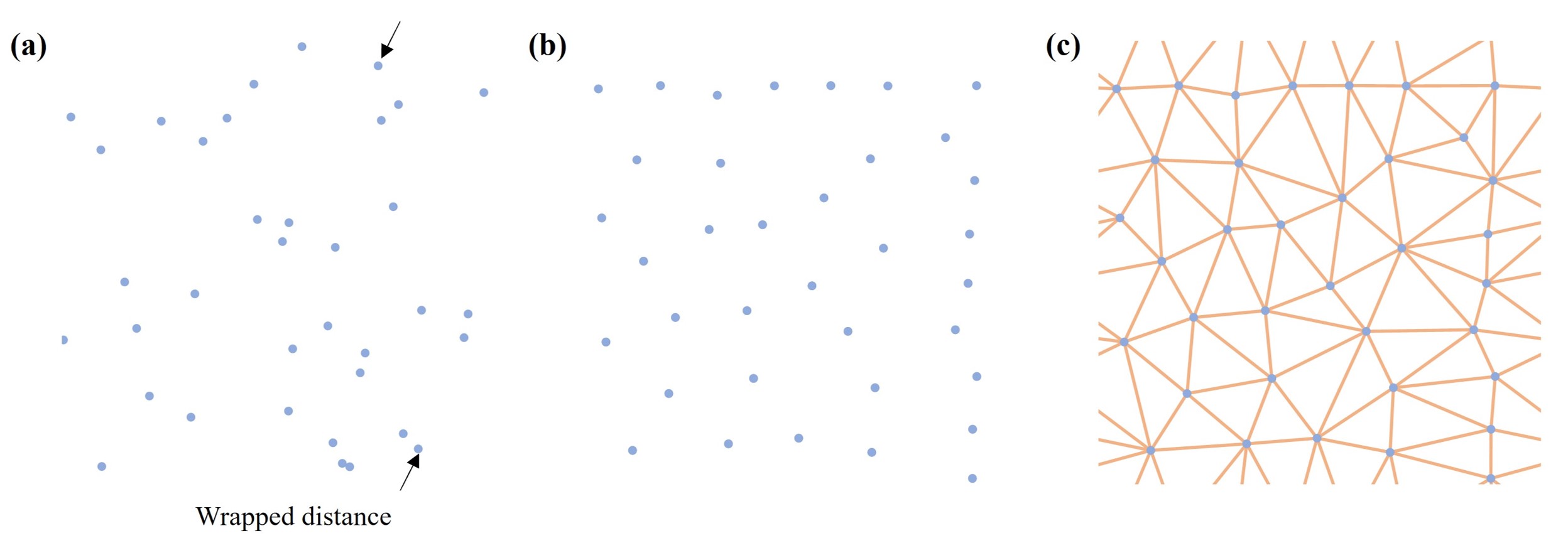}
\caption{\label{sup_fig2} Generating disordered MNN using Delaunay triangulation. (a) N=36 randomly generate points in the range [0,1]. (b) Hyperuniform points with decreased averaged distance and wrapped distance. (c) Delaunay triangulation is formed for all the points in original unit cell and neighboring unit cells. } 
\end{figure}

\begin{figure*}[ht!]
\includegraphics[width=1\linewidth]{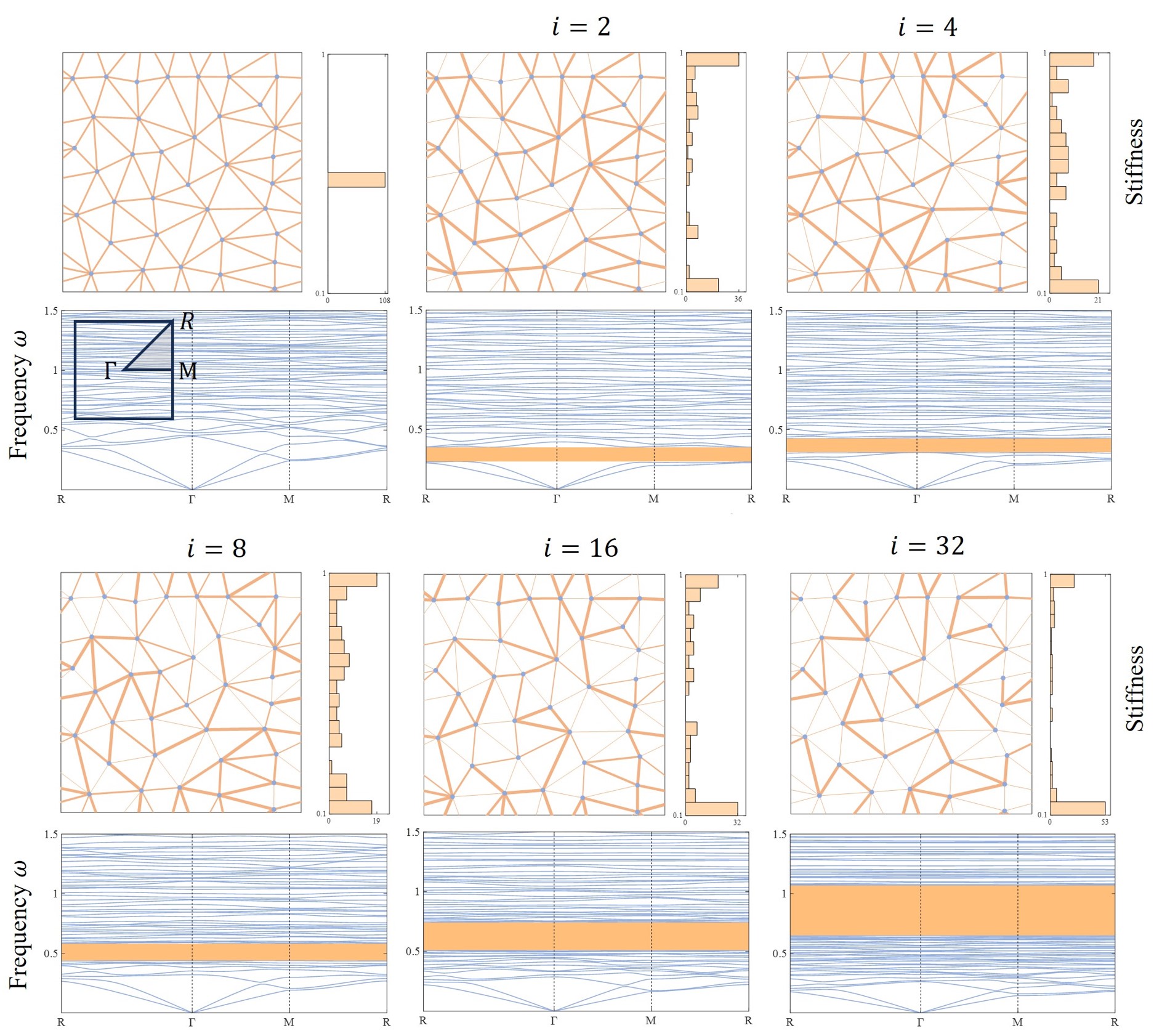}
\caption{\label{fig5} Designing bandgap at the $i$-th band using equilibrium learning. (a-c) The lattice visualization, spring constant distribution, and dispersion curves for (a) untrained lattice, (b-f) lattice trained to open bandgap at (b) 2nd band, (c) 4th band, (d) 8th band, (e) 16th band, and (f) 32nd band. The thickness of the edges is proportioanl to edge stiffness, and the first Brillouin zone is shown in the inset of (a). }
\end{figure*}

\section{Dynamic band control for disordered MNN}
\label{dynamic}
A disordered Delaunay lattice is trained for prescribed dynamical behaviors to demonstrate the versatility and generality of our purposed framework. \cite{ronellenfitsch2019inverse, ronellenfitsch2018optimal} To study the dynamic responses, the MNN consists of hyperuniformly distributed identical masses $m = 1$ and is under Bloch periodic condition in both directions. Learning degrees of freedom $k$, initially set at 0.55, are fine-tuned within the range [0.1, 1].  

The disordered MNN is generated using Delaunay triangulation. Delaunay triangulation is a tessellation method, which tessellate from a given set of points $p_i$ so that no point $p_i$ is inside the circumcircle of any tessellated triangles. First, a set of $N=36$ points are randomly generated within the range [0,1]. An optimization is performed to decrease the average distance or wrapped distance until certain threshold such that the points are hyperuniformly distributed, meaning they are locally random but globally uniform. As shown in Fig. \ref{sup_fig2}(a), wrapped distance refer to points' Euclidean distance from points of neighbor unit cells. The final periodic disordered unit cell is shown in Fig. \ref{sup_fig2}(c).

In order to train the lattice to output target dispersion spectrum, we need to find the gradient of eigenfrequencies \textit{w.r.t.} edge stiffness $\partial \omega/\partial k_i$. The process is similar to that in static MNNs. Let $\lambda = \omega^2$ and mass matrix $M = \mathbf{I}$, the equilibrium in dynamic MNN is given as \cite{lubensky2015phonons}
\begin{equation}
D \pmb{u} - \lambda \pmb{u}=0
\end{equation}
where $D$ denotes the stiffness matrix. Implicit differentiation gives 
\begin{equation}
dD \pmb{u} + D d\pmb{u}- d\lambda \pmb{u} - \lambda d\pmb{u}=0
\end{equation}
Here we normalized eigenvectors to unit length, such that $|\pmb{u}|=1$, and $\pmb{u} \cdot d\pmb{u} = d(|\pmb{u}|)^2=0$. With $\pmb{u}^\dagger$ inserted
\begin{equation}
\pmb{u}^\dagger dD \pmb{u} + \pmb{u}^\dagger D d\pmb{u} - d\lambda=0
\end{equation}
Since $D$ is hermitian, $D^\dagger=D$, and $\pmb{u}^\dagger D = 
 (D \pmb{u})^\dagger = \lambda \pmb{u}^\dagger$
 \begin{equation}
\pmb{u}^\dagger dD \pmb{u} - d\lambda=0
\end{equation}
and eventually, $\partial \omega/\partial k_i$ can be derived from
 \begin{equation}
\pmb{u}^\dagger \frac{\partial D}{\partial k_i} \pmb{u}=\frac{\partial \lambda}{\partial k_i}
\end{equation}

\begin{figure}[ht!]
\includegraphics[width=\linewidth]{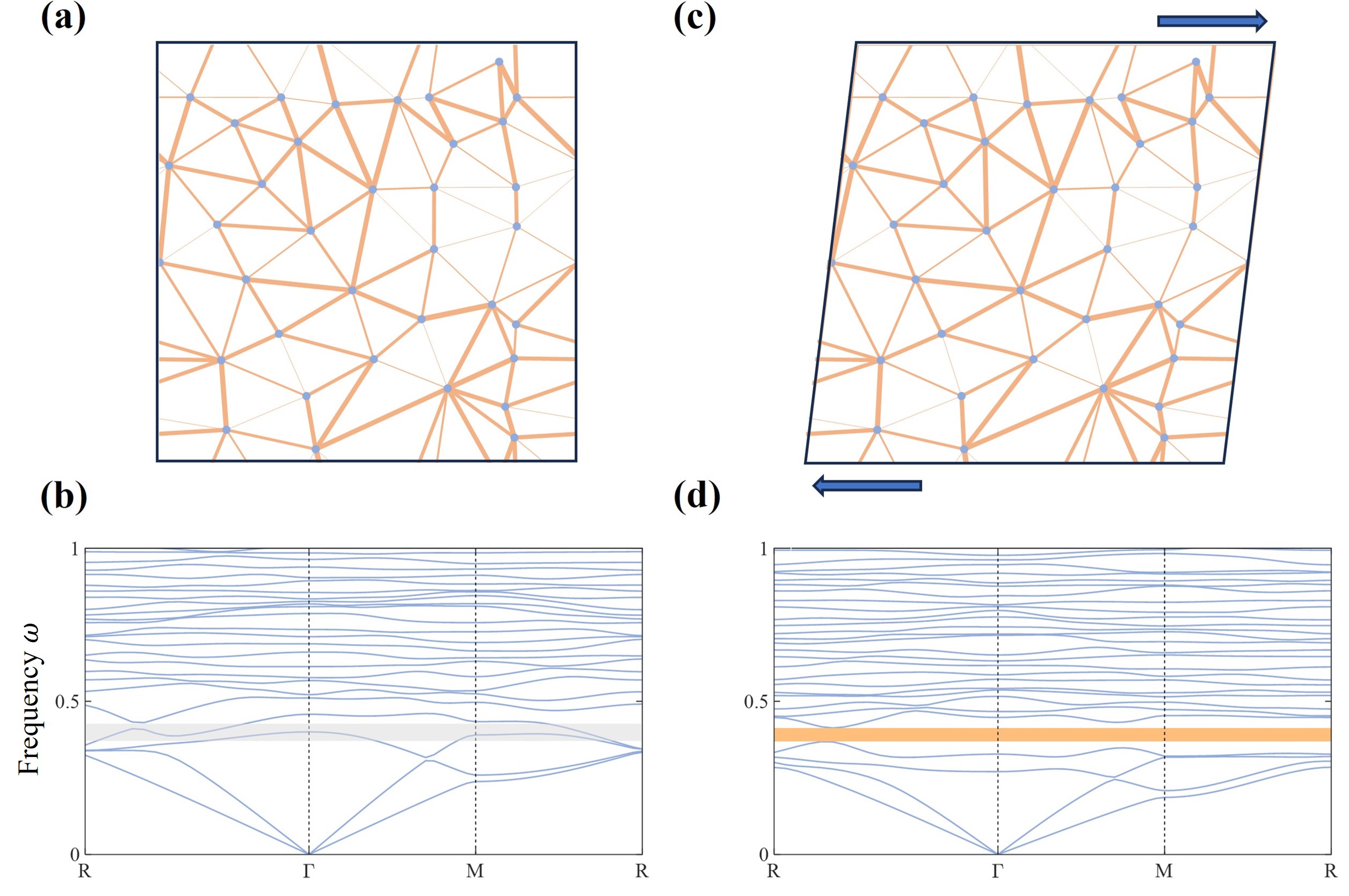}
\centering
\caption{\label{sup_fig3} Mechanical metamaterial as bandgap switch. (a) The trained lattice when no deformation is applied. (b) The dispersion curves for non-deformed lattice. (c) The deformed lattice. (d) The dispersion curves for deformed lattice.} 
\end{figure}

The training loop of a dynamic MNN is shown in bottom panels of Fig. \ref{fig1}. Unlike the static MNN, the dynamic MNN operates with or without input nodes, producing prescribed bandgap sizes as output in the dispersion spectrum. Fig. \ref{fig5} depicts the Delaney lattice, its spring constant distribution, and its dispersion curves before and after training. Specifically, the periodic lattice is trained to open desired bandgaps at the 2nd and 4th bands, resulting in dispersion curves depicted in Fig. \ref{fig5} (b, c). Distinct bandgaps emerge at the specified positions in the spectrum. Analysis of the stiffness distribution reveals a tendency for most edges to be trained toward extremes, exhibiting either very high or very low stiffness. This observation implies a significant impedance mismatch between high and low-stiffness components, leading to bandgaps from Bragg scattering. We open bandgaps at various positions in the spectrum to show the robustness of our back-propagation, as shown in Fig. \ref{fig5} (b-f). 

Beyond the basic tuning of band gaps, the inverse design of metamaterials that adjust their spectral properties on demand in response to an external control stimulus emerges as a significant challenge. Given the nonlinear relationship between the input stimulus, lattice stiffness, and the frequency band, we can train a lattice to adjust its frequency band, $\omega(x,k)$, based on the input stimulus. We can further adapt this framework to design phononic switches. These innovative switches are capable of selectively opening and closing spectral gaps within specific frequency ranges or band indices, according to predefined deformations. The interplay between network response and nonzero edge tensions under strain alters the elastic energy landscape, thereby reshaping the dispersion spectrum. In this context, we have trained the lattice to open a bandgap above the 4th bandgap when it is deformed, and to close this gap in its undeformed state. We shear the lattice in the $x$ direction, with a shear strain of $\epsilon = 0.125$. Figure \ref{sup_fig3} depicts the transformation of bandgaps in response to this deformation. Figure \ref{sup_fig3}(a,b) display the undeformed lattice alongside its frequency spectrum, where no discernible bandgap is apparent in the dispersion relations. Conversely, Figure \ref{sup_fig3}(c,d) showcase the deformed lattice and its corresponding frequency spectrum, revealing a pronounced bandgap centered around a frequency of $\omega = 0.4$ in panel (d). This demonstrates that the presence and location of bandgaps can be dynamically tuned by varying the geometric properties of the periodic lattice. Such capabilities leads to advanced material designs that can be programmed to respond adaptively to external stimuli, opening new avenues in the field of tunable metamaterials.

\section{Conclusion}

In summary, we introduce a back-propagation framework for lattice-based metamaterials built upon the equilibrium within the physical system, and apply it to one- and two-dimensional lattices for desired static and dynamic responses. Such a learning framework allows a lattice to physically learn from external stimulus. By leveraging the implicit differentiation of output displacement with respect to stiffness, we can effectively evaluate updates in each edge and thus train the lattice as an intelligent mechanical neural network to learn various tasks, ranging from shape morphing, and bandgap control to regression. Such MNN has already shown adaptive to manufacturing defects, damages, and material fatigue due to its online learning nature \cite{dillavou2022demonstration}. This framework developed for intelligent mechanical lattices can find important applications in the fast inverse design of various high-dimensional materials. 

The proposed method supports significant flexibility in introducing additional nonlinearities into the system, enhancing its functional capabilities. One approach to achieving this is through the use of nonlinear springs instead of linear ones. Nonlinear stiffness exhibit a non-proportional relationship between force and displacement, allowing for a broader range of responses under varying loads. Additionally, employing a disordered lattice structure provides another avenue for increasing geometric nonlinearity. Unlike orderly and regular lattice structures, disordered lattices lack uniform spacing and alignment, resulting in a more complex interaction of forces within the network. This irregular lattice structure introduces an additional layer of geometric nonlinearity. Our framework can be readily adapted to lattices with nonlinear springs and disordered geometry, potentially significantly enhancing the system's capacity to learn complex behaviors.

\section*{Acknowledgments}
This work is supported by the Air Force Office of Scientific Research under Grant No. AF9550-20-1-0279 with Program Manager Dr. Byung-Lip (Les) Lee.

\section*{Data Availability}
The data that support the findings of this study are available from the corresponding author upon reasonable request

\bibliography{Ref}

\end{document}